\documentclass{article}
\usepackage{amsmath}
\usepackage{amssymb}
\usepackage{waspaa21,amsmath,graphicx,url,times,multirow}
\usepackage{color}
\usepackage{math}

\title{Blind Room Parameter Estimation using Multiple \\ Multichannel Speech Recordings}

\name{Prerak Srivastava,
      Antoine Deleforge,
      Emmanuel Vincent}
\address{Universit\'e de Lorraine, CNRS, Inria, Loria, F-54000 Nancy, France\\ \{prerak.srivastava, antoine.deleforge, emmanuel.vincent\}@inria.fr\\                    
}

\begin{document}

\ninept
\maketitle

\begin{sloppy}

\begin{abstract}

Knowing the geometrical and acoustical parameters of a room may benefit applications such as audio augmented reality, speech dereverberation or audio forensics. In this paper, we study the problem of jointly estimating the total surface area, the volume, as well as the frequency-dependent reverberation time and mean surface absorption of a room in a blind fashion, based on two-channel noisy speech recordings from multiple, unknown source-receiver positions. A novel convolutional neural network architecture leveraging both single- and inter-channel cues is proposed and trained on a large, realistic simulated dataset. Results on both simulated and real data show that using multiple observations in one room significantly reduces estimation errors and variances on all target quantities, and that using two channels helps the estimation of surface and volume. The proposed model outperforms a recently proposed blind volume estimation method on the considered datasets.
\end{abstract}


%

\section{Introduction}
\label{sec:intro}
Audio augmented reality (AAR) has received increased interest in the recent years \cite{valimaki2015assisted}. A key task in AAR is to simulate virtual sound sources that are consistent with the user's environment via, e.g., a hear-through headset.
To achieve this in an immersive way, a major hurdle is to estimate the relevant acoustical parameters of the room the user evolves in. The notion of \textit{reverberation fingerprint} (RF) was introduced in \cite{jot2016augmented} as a compact way to characterize rooms for realistic binaural rendering on AAR headphones. The RF consists of the room's volume $V$ in $\textrm{m}^3$ and its reverberation time per octave band $\textrm{RT}_{60}(b)$ in $\textrm{s}$, where $b\in\mathcal{B}=\{125,250,$ $500,1\textrm{k},2\textrm{k},4\textrm{k}\}~\textrm{Hz}$.
%
%
Under ideal diffuse sound field conditions, these parameters are related via Sabine's well-known formula \cite{eyring1930reverberation}
\begin{equation}
    \textrm{RT}_{60}(b) \approx 0.16\frac{V}{\bar{\alpha}(b)\cdot S}
\end{equation}
where $S=\sum_{i=1}^KS_i$ is the total area of the room's $K$ surfaces in $\textrm{m}^2$ and $\bar{\alpha}(b)=\sum_{i=1}^K\alpha_i(b)S_i/S$ is the area-weighted mean absorption coefficient in octave band $b\in\mathcal{B}$.
%
%
%
%

An attractive research direction to retrieve such parameters is to estimate them solely from audio recordings of unknown sound sources in the room, using microphones placed on the headset. In \cite{murgai2017blind}, an algorithm to blindly estimate the RF based on the decay envelope of a single-channel clean speech signal was proposed and showed encouraging results on simulated data.
In the same year, the authors of \cite{kataria2017hearing} trained a statistical model on a large dataset generated from simulated room impulse responses (RIRs) to blindly estimate the position of a broadband source and the mean absorption coefficient of walls above 1~kHz using a binaural receiver and inter-channel cues. Experiments were limited to a fixed room geometry, a fixed receiver position, and simulated data. Later, \cite{microsoft_vol} paved the way for blind room volume estimation from single-channel noisy speech using a convolutional neural network trained using both simulated and real RIRs. The reported results showed that this method can estimate a broad range of volumes within a factor of 2 on real data from the Acoustic Characterization of Environments (ACE) challenge \cite{ace2016james}.
Following the ACE challenge, a number of neural-network-based methods were proposed to blindly estimate reverberation times and direct-to-reverberant ratios from speech signals \cite{DBLP:journals/corr/XiongGM15, microsoft_rt60, chai2019blind, impulse2020bryan}. In particular, \cite{impulse2020bryan} proposes a method to generate augmented training datasets from real RIRs. These works estimate broadband values rather than frequency-dependent ones, which may limit the rendering realism in the context of AAR. They also all use single-channel signals. 
%
As an interesting alternative approach, \cite{su2020acoustic} proposes to learn a latent representation referred to as \textit{room embedding}, which is then used to condition a waveform-to-waveform network that converts a signal recorded in one environment to another. However, this approach cannot be straightforwardly applied to spatial AAR, whereas explicitly knowing room parameters offers more flexibility in binaural rendering. Moreover, explicit room parameters may be useful in themselves for other applications such as speech dereverberation \cite{wu2017reverberation} or audio forensics \cite{moore2013roomprints}.

In this paper, we push these recent research efforts further by proposing a blind method to jointly estimate the total surface area $S$, the volume $V$, the reverberation times $\textrm{RT}_{60}(b)$ and the mean absorption coefficients $\bar{\alpha}(b)$ in all octave bands $b\in\mathcal{B}$. On top of that, we investigate whether using multichannel input and/or fusing estimates across multiple source-receiver positions inside the same room help. To this end, we propose a new convolutional neural network architecture that combines single- and inter-channel features in a joint embedding layer, and uses a likelihood-based loss function that yields adaptive variance estimates, which allow us to fuse multiple independent observations of the same room in a statistically principled way. The model is trained on a carefully-crafted realistic simulated dataset and tested on both simulated and real wet speech recordings\footnote{Our code to reproduce this paper is available at \url{github.com/prerak23/RoomParamEstim}.}. 


\begin{figure*}[t!]
  \centering
  \centerline{\includegraphics[width=\linewidth]{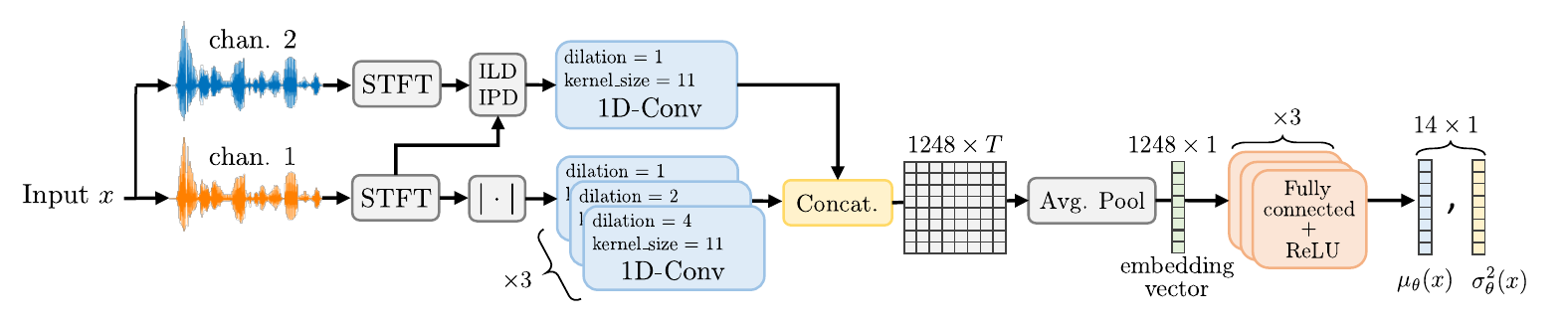}}
  \caption{Diagram of the proposed neural network architecture.}
  \label{fig:diagram}
\end{figure*}

\section{Simulated Dataset}
\label{sec:data}
To train and validate our approach, a large dataset of noisy speech signals annotated with target room acoustic parameters is created. First, synthetic RIRs are generated using 
Roomsim \cite{roomsim}. 
Roomsim is a hybrid shoe-box room acoustic simulator combining the image-source method \cite{imagesrc} to simulate specular reflections that dominate the early part of RIRs, and the diffuse-rain method \cite{roomsim} to model scattered reflections using stochastic ray-tracing. Simulations are run using a sampling frequency of 48~kHz, a reflection order of 10 for the image-source method and 2000 rays for the diffuse-rain method. 20,000 rooms are simulated with length, width and height drawn uniformly at random in $[3, 10]$~m, $[3, 10]$~m and $[2.5, 4]$~m, respectively, resulting in $S\in[48,360]$ $\textrm{m}^2$ and $V\in[18,400]$ $\textrm{m}^3$. For each room, 5 RIRs corresponding to different source-receiver positions are generated. Omnidirectional sources and receivers are used and placed uniformly at random at least 30~cm from each surface and from each other. The receiver is a two-microphone array placed parallel to the floor with an aperture of 22.5~cm, that is a typical head or headset width.
For each room, a single scattering coefficient drawn uniformly at random in $[0.2,1]$ is used on all surfaces and octave bands for the diffuse-rain method. The absorption coefficients of the 6 surfaces in 6 octave bands are sampled using the \textit{reflectivity-biased} sampling strategy proposed in \cite{cedricfoy}. Each $\alpha_i(b)$ is drawn uniformly at random inside ranges based on measured databases of commonly encountered surface materials, while giving each surface a 50$\%$ probability of having a frequency-independent reflective profile ($\alpha_i(b)<0.12$). As showed in \cite{cedricfoy}, this yields more realistic and more diverse distributions of absorption and reverberation times ($\textrm{RT}_{60}(b)\in[0.2,3.2]$~s and $\bar{\alpha}(b)\in[0.02,0.6]$ in our dataset) than, e.g., by sampling each $\alpha_i(b)$ uniformly in $[0, 1]$.

The obtained RIRs are downsampled to 16~kHz and convolved with random speech excerpts from the LibriSpeech \cite{librispeech} corpus. The resulting 3~s two-channel reverberated signals are then corrupted with both static microphone noise, i.e., independent additive white Gaussian noise on each channel, and spatially-diffuse babble noise, i.e., speech-shaped noise convolved with the late part ($>$50~ms) of a random RIR in the room. For the noise levels to be realistic, signals from sources that are placed further away from the receiver should exhibit lower signal-to-noise ratios (SNRs).
To achieve this, for each room, we first generate a \textit{reference} signal using a random speech source placed 1 meter in front of a receiver (not used in the final dataset). Static and diffuse noise levels for this signal are set to obtain SNRs drawn uniformly at random in $[70,90]~\textrm{dB}$ and $[30,60]~\textrm{dB}$, respectively (different levels for each source-receiver position in the room are set). These noise levels are then kept fixed for the final mixtures, irrespective of the distance from the speech source to the receiver. This resulted in an overall SNR range of $[-10,65]~\textrm{dB}$ across the dataset.

The 5$\times$20k = 100k two-channel noisy speech signals are divided into training, validation and test sets of respective size 80k, 10k and 10k with no room or speech-signal overlap between them. Each room is annotated with the 14 target parameters, namely, 6 mean absorption coefficients and 6 reverberation times in all octave bands, the total surface area $S$ and the volume $V$. For each room and each octave band, a unique reverberation time is estimated by taking the median value over the 5 source-receiver positions available. The values are obtained by linear regression over the -5~dB to -25~dB decay of Schroeder curves \cite{schroeder1965new}.



\section{Neural Network Model}
\label{sec:neural_net}
\subsection{Proposed Architecture, Cost Function, and Fusion Method}
The proposed neural network architecture is depicted in Fig.~\ref{fig:diagram}. 
Single-channel (SC) and inter-channel (IC) features are extracted from the time-domain two-channel input signal $\xvect$ in the form of spectrograms. We use the short-time Fourier transform (STFT) with 96~ms Hann windows and 50\% overlap to obtain a complex spectrogram $\{X_i(f,t)\}_{f,t=1}^{F,T}$ for each channel $i$, with $F=769$ positive frequency bins and $T=63$ time frames for a 3 s input signal (our architecture works on arbitrary input length). This choice performed best among 32, 64, 96 and 128~ms windows in our preliminary experiments. Then, SC features are computed as $|X_1(f,t)|$, which performed better than $|X_1(f,t)|^2$, $|X_1(f,t)|^{1/2}$, or $\log|X_1(f,t)|$. IC features are obtained by concatenating inter-channel level differences (ILD) and phase differences (IPD):
\begin{align}
    \textrm{ILD}(f,t) &= \log |X_1(f,t)| - \log |X_2(f,t)|, \\
    \textrm{IPD}(f,t) &= \left[\operatorname{Re},\operatorname{Im} \left(\frac{X_1(f,t) X_2^{*}(f,t)}{|X_1(f,t) X_2^{*}(f,t)|}\right)  \right].
\end{align}
These features are then processed through 1D convolutional blocks (1D-Conv), which were recently proposed in the Conv-TasNet architecture in the context of speech separation \cite{convtasnet}. These blocks consist of separable convolutions (depth-wise and point-wise) intertwined with rectified linear unit (ReLU) activations and followed by layer normalization \cite{layernorm}. The latter proved to be crucial in our experiments, as it creates scale-invariant representations. For SC features, three 1D-Conv blocks with increasing dilation factors along the frequency-axis and a kernel size of 11 are used, while only one block is used on IC features, as this showed to give best results. The obtained representations are concatenated along the frequency axis and average-pooled along the time-axis to yield a time-independent, 1248-dimensional \textit{embedding vector}. The embedding vector is finally passed through 3 fully-connected layers of respective dimensions 96, 48 and 28 to obtain $2\times D=14$ outputs consisting of the estimated room parameters $\muvect_{\boldsymbol{\theta}}(\xvect)\in\mathbb{R}^{D}$  and the estimated variances $\sigmavect^2_{\boldsymbol{\theta}}(\xvect)\in\mathbb{R}^{D}$ (or uncertainties) on these parameters.


The network parameters $\boldsymbol{\theta}$ are optimized by minimizing the following Gaussian negative log-likelihood loss function:
\begin{align}
    \mathcal{L}_{\boldsymbol{\theta}}(\xvect,\yvect)&=-\log p_{\boldsymbol{\theta}}(\yvect|\xvect)= -\log\mathcal{N}(\yvect;\muvect_{\boldsymbol{\theta}}(\xvect),\sigmavect^2_{\boldsymbol{\theta}}(\xvect)) \nonumber \\
    \label{eq:ML_loss}
    &\stackrel{c}{=} \frac{1}{2}\sum_{d=1}^{D}\log\sigma^2_{d,\boldsymbol{\theta}}(\xvect)
                     +\frac{(y_d-\mu_{d,\boldsymbol{\theta}}(\xvect))^2}{\sigma_{d,\boldsymbol{\theta}}^2(\xvect)}
\end{align}
where $\yvect\in\mathbb{R}^{D}$ denotes the ground truth room parameters. 
A benefit of this approach is that it adaptively weights errors on individual parameters. 
The estimated variances can also be used to fuse estimated obtained from $J$ independent observations $\{\xvect_j\}_{j=1}^J$ of the same room using the following formula derived from Bayes' rule:
\begin{gather}
\label{eq:multi_input}
    p_{\boldsymbol{\theta}}(y_d|\bar{\xvect}=[\xvect_1,\dots,\xvect_J])=\mathcal{N}(y_d;\bar{\mu}_{d,\boldsymbol{\theta}}(\bar{\xvect}),1/\bar{\gamma}_{d,\boldsymbol{\theta}}^{2}(\bar{\xvect})) \\
    \text{with }\bar{\mu}_{d,\boldsymbol{\theta}}(\bar{\xvect})=\sum_{j=1}^J\frac{\gamma_{d,\boldsymbol{\theta}}^2(\xvect_j)}{\bar{\gamma}_{d,\boldsymbol{\theta}}^2(\bar{\xvect})}\mu_{d,\boldsymbol{\theta}}(\xvect_j),\;
    \bar{\gamma}_{d,\boldsymbol{\theta}}^2(\bar{\xvect}) = \sum_{j=1}^J\gamma_{d,\boldsymbol{\theta}}^2(\xvect_j) \nonumber
\end{gather}
where $\gamma_{d,\boldsymbol{\theta}}^2(\xvect_j) = 1/\sigma_{d,\boldsymbol{\theta}}^2(\xvect_j)$ is the estimated precision for observation $\xvect_j$ and $\bar{\muvect}_{d,\boldsymbol{\theta}}(\bar{\xvect})$ is the fused estimate.

To avoid issues due to scale differences, the ground truth parameters are normalized at training time by dividing them by their standard deviations over the training set, which are saved and multiplied with the network output at test time.
The network is trained using the ADAM optimizer \cite{kingma2014adam} with a learning rate of $10^{-4}$ and a batch size of 120. A dropout rate of 0.2 and 0.4 was used in conv-blocks and in fully connected layers to avoid over-fitting. We used a patience of 15 epochs on the validation set for early stopping. Our models generally converged in 100--150 epochs.   

\subsection{Other Tried Architectures}
In addition to 1D-Conv blocks, Conv-TasNet introduced the idea of learnable input filters \cite{convtasnet}. We tried replacing the STFT by such learnable filters to estimate our SC features, but this yielded worse performance. We also tried an architecture inspired by x-vectors, a state-of-the-art model for speaker recognition \cite{snyder2018x}, but this did not perform well on our data. We experimented with replacing the proposed maximum-likelihood loss (\ref{eq:ML_loss}) by a simple mean squared error loss. This yielded significantly higher errors while losing the benefit of easily fusing multiple observations in one room. Estimating $\log(V)$ and $\log(S)$ instead of $V$ and $S$ as suggested in \cite{microsoft_vol} did not improve results on our data, most likely because the ranges we consider for these quantities are smaller. Finally, instead of estimating $\bar{\alphavect}, \textrm{RT}_{60}, V$ and $S$ jointly in a multi-task fashion, we trained the same architecture to estimate each of these quantities individually. The obtained results were identical, while increasing the number of model parameters and training time by a factor of 4, which suggests that our proposed network achieves efficient weight sharing.

\section{Experiments and results}
\label{sec:expe}
\subsection{Simulated Data}
\begin{figure}[t]
  \centering
  \centerline{\includegraphics[width=\columnwidth]{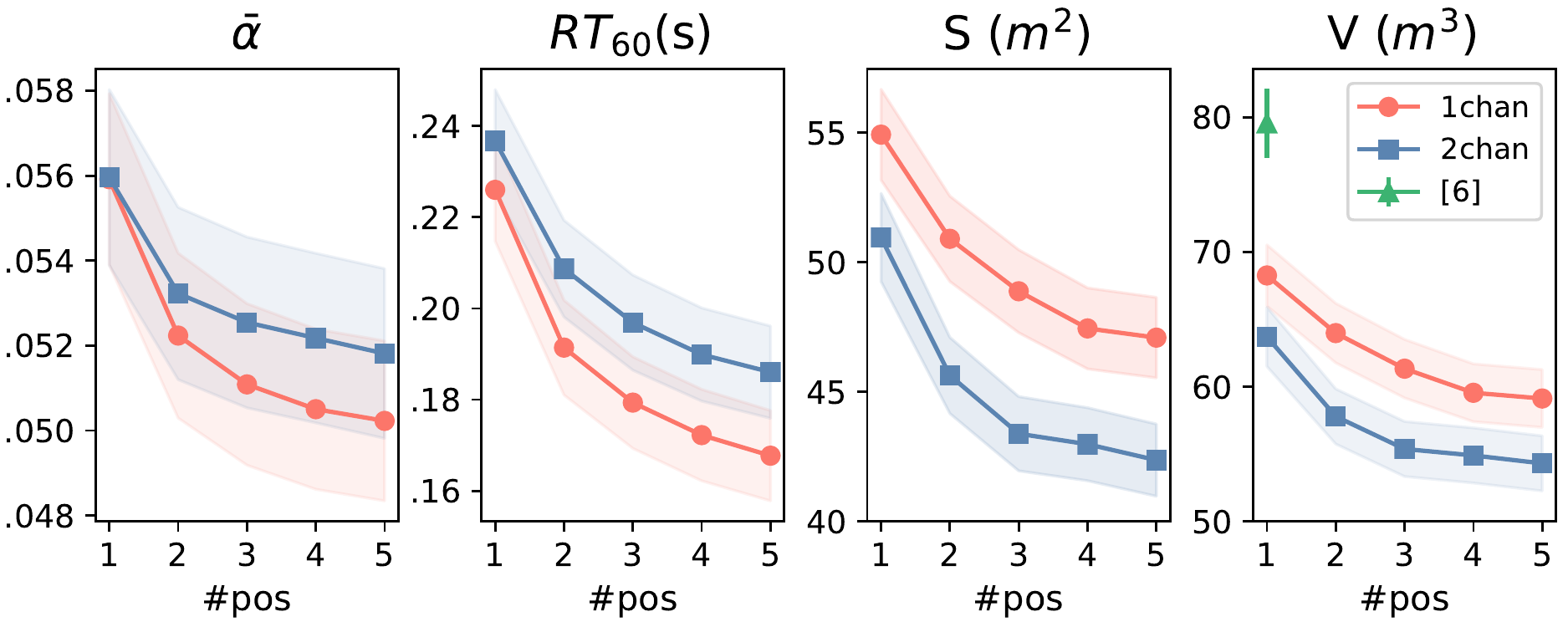}}
  \caption{Mean absolute error achieved on simulated data by \cite{microsoft_vol} vs.\ the proposed model with one- or two-channel inputs, as a function of the number of source-receiver positions fused in each room. Shaded areas indicate 95\% confidence intervals.}
  \label{fig:fusion_vps}
\end{figure}

\begin{table}[t]
    \centering
    \footnotesize
    \begin{tabular}{ c|c|c|c|c|c}
\hline
Input & Feature & $\bar{\alpha}$ & $\textrm{RT}_{60}~(\textrm{s})$ & $S~(\textrm{m}^2)$ & $V~(\textrm{m}^3)$ \\
\hline
1mic, 1sig & SC & 0.055 & 0.225 & 55.0 & 68.8\\ 
1mic, 2sig & SC & 0.058 & 0.222 & 55.4 & 69.7  \\ 
2mic, 1sig & SC & 0.057 & 0.221 & 54.0 & 67.7 \\
2mic, 1sig & SC+IC & 0.055 & 0.236 & \textbf{49.9} & \textbf{63.7}\\
\hline
\end{tabular}
    \caption{Mean absolute error achieved on simulated data for one source-receiver position using different inputs and features. Bold numbers indicate the best statistically significant result per column based on 95\% confidence intervals, when there is one.}
    \label{tab:SC-IC}
\end{table}

\begin{table}[t]
    \centering
    \footnotesize
    \begin{tabular}{ c|c|c|c|c}
\hline
\multirow{2}{*}{Octave bands} & \multicolumn{2}{c}{1 position}     & \multicolumn{2}{|c}{5 positions}   \\
\cline{2-5}
 & $\bar{\alpha} $ & $\textrm{RT}_{60}~(\textrm{s})$  & $\bar{\alpha} $ & $\textrm{RT}_{60}~(\textrm{s})$ \\
\hline
125~Hz & 0.056 & 0.392 & 0.051 & 0.320 \\ 
250~Hz & 0.060 & 0.305 & 0.055 & 0.249   \\
500~Hz & 0.057 & 0.228 &  0.051 & 0.170  \\
1~kHz & 0.053 & 0.188 & 0.050 & 0.146   \\
2~kHz & 0.054 & 0.165 & 0.051 & 0.122   \\
4~kHz & 0.052 & 0.139 & 0.049 & 0.106   \\
\hline
\end{tabular}
    \caption{Mean absolute errors on $\bar{\alpha}(b)$ and $\textrm{RT}_{60}(b)$ in the 6 octave bands achieved by the proposed model with two channels and 1 or 5 source-receiver positions per room on simulated data.}
    \label{tab:octave_results}
\end{table}
We first evaluate the proposed approach on the 2,000 unseen rooms of the simulated test set (see Section \ref{sec:data}), using between 1 and 5 two-channel 3-s noisy speech recordings at different source-receiver positions per room, based on the fusion method in (\ref{eq:multi_input}). 
Two variants of our model are compared: the full two-channel architecture depicted in Fig.~\ref{fig:diagram}, and the same architecture without the upper inter-channel processing part of the network, i.e., using only one channel. These variants are compared to our implementation of the single-channel, single-position blind room volume estimation method of \cite{microsoft_vol} trained on the same data. We could not find competing methods for blind total surface estimation or blind frequency-dependent reverberation time and mean absorption coefficient estimation in the literature. The metric used is the mean absolute error on each room parameter.



As can be seen in Fig.~\ref{fig:fusion_vps}, increasing the number of fused observations per room significantly reduces errors on all parameters. Using five positions and two channels, mean absolute errors of $0.052$ for $\bar{\alpha}$ (training range $[0.02,0.6]$), $0.18~\textrm{s}$ for $\textrm{RT}_{60}$ (training range $[0.2, 3.2]$), $42~\textrm{m}^2$ for $S$ (training range $[48,360]$) and $54~\textrm{m}^3$ for $V$ (training range $[18,400]$) are obtained. The proposed model significantly outperforms the one in \cite{microsoft_vol} for volume estimation, reducing the error by 13\% using one channel and one observation, and by 31\% using two channels and five observations. Interestingly, we observe that using two channels instead of one significantly reduces surface and volume estimation error without significantly impacting mean absorption and reverberation time estimation, according to 95\% confidence intervals. This may be interpreted by the fact that the latter mostly govern the late, spatially-diffuse regime of RIRs, and hence should have limited correlation with inter-channel cues that mostly capture spatial characteristics. Conversely, $S$ and $V$ are inherently spatial quantities as they relate to the room's geometry and hence early echoes, which do correlate with IC cues \cite{di2019mirage, dechorate}.

We hence carry an additional experiment to study whether the improvement observed when using two channels is truly explained by concatenating SC and IC representations at the embedding layer (see Fig.~\ref{fig:diagram}), or is only due to doubling the number of signal samples in a given source-receiver position, which mitigates noise. Four different embeddings are compared, namely (i) the SC representation of one speech signal (1mic, 1sig), (ii) the average of the SC representations of two speech signals (1mic, 2sig), (iii) the concatenation of the SC representations of the two individual channels of one speech signal (2mic, 1sig) and (iv) the proposed concatenation of SC and IC representations. Speech signals are always 3-s long.
As can be seen in Table~\ref{tab:SC-IC}, the only model which significantly improves surface and volume estimation is the one using IC features, despite the fact that the 2nd and 4th models also benefit from the same number of input signal samples.

So far, the reported errors on $\bar{\alpha}$ and $\textrm{RT}_{60}$ were averaged over all 6 octave bands. Table~\ref{tab:octave_results} reports detailed errors per octave band. As can be seen, the mean absorption is well estimated in all octave bands, while reverberation-time errors steadily decrease by a factor of 3 from 125~Hz to 4~kHz. This could be explained by the fact that less information is available in the narrower, lower octave bands. 

\subsection{Real Data}

To check how well our method generalizes to real data, we use the recently released dEchorate dataset \cite{dechorate}, which contains wet speech recordings made inside the acoustic lab at Bar-Ilan University. The lab is a shoe-box room of size $5.7\times6\times 2.4~\textrm{m}$ ($S=125~\textrm{m}^2$, $V=82~\textrm{m}^2$) where each of the walls, floor and ceiling can be set either to either a reflective mode or an absorbent mode. 5 arrays of 6 omnidirectional microphones and 6 directional loudspeakers are placed inside the room, yielding $5\times 6=30$ multi-channel speech recordings per room configuration. For each room, ground truth $\bar{\alpha}$ and $\textrm{RT}_{60}$ values are provided for the four octave bands from 500~Hz to 4~kHz. Hence, results in lower octave bands are omitted here. In our experiments, we use $3\times 30=90$ 3-s speech recordings corresponding to the 2-channel sub-arrays with aperture $22.5~\textrm{cm}$ and to the 3 room configurations involving 3 or more reflective surfaces, as these most closely match the considered scenario. For these rooms, $\bar{\alpha}(b)$ ranges from 0.16 to 0.35 and $\textrm{RT}_{60}(b)$ from 0.25 to 0.66~s.


Table \ref{tab:real_results} reports mean absolute errors using the proposed approach with or without IC features. We report results using either 1 or 5 source positions and a fixed receiver. For the latter, we exclude one out of the 6 available source positions for each test, so that there are 90 tests in each case. The mean absolute volume estimation error using the single-channel, single-position approach of \cite{microsoft_vol} is reported as well. Encouragingly, errors obtained with our approach are of comparable orders to those obtained on simulated data. In the single-channel, single-position case, volume estimation errors obtained with our model are comparable to \cite{microsoft_vol}. For $\textrm{RT}_{60}$, $S$ and $V$, we are able to reproduce the observation that increasing the number of source-receiver positions significantly decreases errors, at least using IC features. We also observe again that using IC features significantly improves the estimation of $S$ and $V$. The errors obtained on $\bar{\alpha}$ are less consistent than those on simulated data, which may be due to the difficulty of reliably annotating mean absorption coefficients in real rooms.
\begin{table}[t]
    \centering
    \footnotesize
    \begin{tabular}{c|c|c|c|c|c|c}
\hline
Method & Features & \# pos & $\bar{\alpha} $ & $\textrm{RT}_{60}$ & $S$ & $V$ \\
\hline
\cite{microsoft_vol} & SC & 1  & - & - & - & 137.8 \\
Ours & SC & 1  & \textbf{0.061} &0.134 & 129.6 & 154.5 \\ 
Ours & SC & 5  & \textbf{0.060} &0.097& 125.8 & 149.1 \\
Ours & SC+IC & 1 & 0.084 &0.101 & 89.4 & 107.6 \\ 
Ours & SC+IC & 5 & 0.094& \textbf{0.062}& \textbf{50.2} & \textbf{68.8}\\
\hline
\end{tabular}
    \caption{Mean absolute error achieved over 3 rooms from the real dEchorate dataset. Bold numbers indicate the best statistically significant result per column, based on 95\% confidence intervals.}
    \label{tab:real_results}
\end{table}
\begin{table}[t]
    \centering
    \footnotesize
    \begin{tabular}{c|c|c|c|c|c|c}
\hline
Method & Features & \# pos & $\bar{\alpha} $ & $\textrm{RT}_{60}$ & $S$ & $V$ \\
\hline
\cite{microsoft_vol} & SC & 1  & - & - & - & 10.0 \\
Ours & SC & 1  & 0.030 & 0.161 & 27.2 & 31.8\\ 
Ours & SC & 5  & 0.024 & 0.090 & 19.6 & 23.0 \\
Ours & SC+IC & 1 & 0.031 &0.100 & 34.7 & 39.7 \\ 
Ours & SC+IC & 5 & 0.015& 0.054 & 16.5 & 18.9\\
\hline
\end{tabular}
    \caption{Standard deviation of parameter estimates for room "011100" of the real dEchorate dataset.}
    \label{tab:real_results_std}
\end{table}
%

Finally, Table \ref{tab:real_results_std} shows the standard deviations of estimated values by the same methods over the 30 recordings from the room with 3 reflective surfaces. Encouragingly, the relatively low standard deviations reveal the ability of the models to provide parameter estimates that are stable within a room, and do not depend much on the source-receiver position. Moreover, it can be seen that, as expected, using five observations in a room instead of one systematically decreases the standard deviation of estimates. 

\section{Conclusion}
\label{sec:concl}
%
This study revealed that using inter-channel cues can significantly improve the blind estimation of a room's volume and surface from noisy speech, while for estimating reverberation and absorption parameters, a single channel is sufficient. It also highlights that fusing multiple measurements reduces estimation errors and variances on all parameters. Finally, we showed that a system trained on a carefully simulated training set offers reasonable generalization capabilities to real data. Future work will include the use of data-augmentation and domain adaptation techniques to improve real data results, extensions to binaural or ambisonic receivers, and the joint estimation of local parameters such as the positions and properties of the source and individual surfaces in the room, with the help of early acoustic echoes.

\section{Acknowledgments}
This work was made with the support of the French National Research Agency through project HAIKUS ``Artifical Intelligence applied to augmented acoustic scenes'' (ANR-19-CE23-0023). Experiments presented in this paper were carried out using the Grid’5000 testbed, supported by a scientific interest group hosted by Inria and including CNRS, RENATER and several Universities as well as other organizations (see https://www.grid5000.fr). We thank the authors of \cite{microsoft_vol} for their helpful advice.

\bibliographystyle{IEEEtran}
\bibliography{refs21}

\begin{thebibliography}{10}
\providecommand{\url}[1]{#1}
\def\UrlFont{\rmfamily}
\providecommand{\newblock}{\relax}
\providecommand{\bibinfo}[2]{#2}
\providecommand\BIBentrySTDinterwordspacing{\spaceskip=0pt\relax}
\providecommand\BIBentryALTinterwordstretchfactor{4}
\providecommand\BIBentryALTinterwordspacing{\spaceskip=\fontdimen2\font plus
\BIBentryALTinterwordstretchfactor\fontdimen3\font minus
  \fontdimen4\font\relax}
\providecommand\BIBforeignlanguage[2]{{%
\expandafter\ifx\csname l@#1\endcsname\relax
\typeout{** WARNING: IEEEtran.bst: No hyphenation pattern has been}%
\typeout{** loaded for the language `#1'. Using the pattern for}%
\typeout{** the default language instead.}%
\else
\language=\csname l@#1\endcsname
\fi
#2}}

\bibitem{valimaki2015assisted}
V.~Valimaki, A.~Franck, J.~Ramo, H.~Gamper, and L.~Savioja, ``Assisted
  listening using a headset: Enhancing audio perception in real, augmented, and
  virtual environments,'' \emph{IEEE Signal Processing Magazine}, vol.~32,
  no.~2, pp. 92--99, 2015.

\bibitem{jot2016augmented}
J.-M. Jot and K.~S. Lee, ``Augmented reality headphone environment rendering,''
  in \emph{2016 AES International Conference on Audio for Virtual and Augmented
  Reality}, 2016, pp. 8--2.

\bibitem{eyring1930reverberation}
C.~F. Eyring, ``Reverberation time in “dead” rooms,'' \emph{The Journal of
  the Acoustical Society of America}, vol.~1, no.~2A, pp. 217--241, 1930.

\bibitem{murgai2017blind}
P.~Murgai, M.~Rau, and J.-M. Jot, ``Blind estimation of the reverberation
  fingerprint of unknown acoustic environments,'' in \emph{AES 143rd
  Convention}, 2017, p. 9905.

\bibitem{kataria2017hearing}
S.~Kataria, C.~Gaultier, and A.~Deleforge, ``Hearing in a shoe-box: binaural
  source position and wall absorption estimation using virtually supervised
  learning,'' in \emph{2017 IEEE International Conference on Acoustics, Speech
  and Signal Processing (ICASSP)}, 2017, pp. 226--230.

\bibitem{microsoft_vol}
A.~F. Genovese, H.~Gamper, V.~Pulkki, N.~Raghuvanshi, and I.~J. Tashev, ``Blind
  room volume estimation from single-channel noisy speech,'' in \emph{2019 IEEE
  International Conference on Acoustics, Speech and Signal Processing
  (ICASSP)}, 2019, pp. 231--235.

\bibitem{ace2016james}
J.~Eaton, N.~D. Gaubitch, A.~H. Moore, and P.~A. Naylor, ``Estimation of room
  acoustic parameters: The {ACE} challenge,'' \emph{IEEE/ACM Transactions on
  Audio, Speech, and Language Processing}, vol.~24, no.~10, pp. 1681--1693,
  2016.

\bibitem{DBLP:journals/corr/XiongGM15}
F.~Xiong, S.~Goetze, and B.~T. Meyer, ``Joint estimation of reverberation time
  and direct-to-reverberation ratio from speech using auditory-inspired
  features,'' \emph{arXiv preprint arXiv:1510.04620}, 2015.

\bibitem{microsoft_rt60}
H.~Gamper and I.~J. Tashev, ``Blind reverberation time estimation using a
  convolutional neural network,'' in \emph{16th International Workshop on
  Acoustic Signal Enhancement (IWAENC)}, 2018, pp. 136--140.

\bibitem{chai2019blind}
M.~Chai, T.~Li, M.~Zhu, T.~Wang, and W.~Zhang, ``Blind estimation of
  reverberation time using binaural complex ideal ratio mask,'' in \emph{2019
  IEEE International Conference on Multimedia \& Expo Workshops (ICMEW)}, 2019,
  pp. 378--383.

\bibitem{impulse2020bryan}
N.~J. Bryan, ``Impulse response data augmentation and deep neural networks for
  blind room acoustic parameter estimation,'' in \emph{2020 IEEE International
  Conference on Acoustics, Speech and Signal Processing (ICASSP)}, 2020, pp.
  1--5.

\bibitem{su2020acoustic}
J.~Su, Z.~Jin, and A.~Finkelstein, ``Acoustic matching by embedding impulse
  responses,'' in \emph{2020 IEEE International Conference on Acoustics, Speech
  and Signal Processing (ICASSP)}, 2020, pp. 426--430.

\bibitem{wu2017reverberation}
B.~Wu, M.~Yang, K.~Li, Z.~Huang, S.~M. Siniscalchi, T.~Wang, and C.-H. Lee, ``A
  reverberation-time-aware {DNN} approach leveraging spatial information for
  microphone array dereverberation,'' \emph{EURASIP Journal on Advances in
  Signal Processing}, vol. 2017, no.~1, pp. 1--13, 2017.

\bibitem{moore2013roomprints}
A.~H. Moore, M.~Brookes, and P.~A. Naylor, ``Roomprints for forensic audio
  applications,'' in \emph{2013 IEEE Workshop on Applications of Signal
  Processing to Audio and Acoustics (WASPAA)}, 2013, pp. 1--4.

\bibitem{roomsim}
S.~M. Schimmel, M.~F. Muller, and N.~Dillier, ``A fast and accurate
  “shoebox” room acoustics simulator,'' in \emph{2009 IEEE International
  Conference on Acoustics, Speech and Signal Processing (ICASSP)}, 2009, pp.
  241--244.

\bibitem{imagesrc}
J.~Allen and D.~Berkley, ``Image method for efficiently simulating small-room
  acoustics,'' \emph{The Journal of the Acoustical Society of America},
  vol.~65, pp. 943--950, 1979.

\bibitem{cedricfoy}
C.~Bastien, A.~Deleforge, and C.~Foy, ``Mean absorption coefficient estimation
  from impulse responses: Deep learning vs.\ {Sabine},'' in \emph{E-FA 2020 -
  Forum Acusticum 2020}, Dec. 2020.

\bibitem{librispeech}
V.~Panayotov, G.~Chen, D.~Povey, and S.~Khudanpur, ``Librispeech: An {ASR}
  corpus based on public domain audio books,'' in \emph{2015 IEEE International
  Conference on Acoustics, Speech and Signal Processing (ICASSP)}, 2015, pp.
  5206--5210.

\bibitem{schroeder1965new}
M.~R. Schroeder, ``New method of measuring reverberation time,'' \emph{The
  Journal of the Acoustical Society of America}, vol.~37, no.~6, pp.
  1187--1188, 1965.

\bibitem{convtasnet}
Y.~Luo and N.~Mesgarani, ``{Conv-TasNet}: Surpassing ideal time–frequency
  magnitude masking for speech separation,'' \emph{IEEE/ACM Transactions on
  Audio, Speech, and Language Processing}, vol.~27, no.~8, p. 1256–1266, Aug
  2019.

\bibitem{layernorm}
J.~L. Ba, J.~R. Kiros, and G.~E. Hinton, ``Layer normalization,'' \emph{stat},
  vol. 1050, p.~21, 2016.

\bibitem{kingma2014adam}
\BIBentryALTinterwordspacing
D.~P. Kingma and J.~Ba, ``Adam: A method for stochastic optimization,'' in
  \emph{ICLR (Poster)}, 2015. [Online]. Available:
  \url{http://arxiv.org/abs/1412.6980}
\BIBentrySTDinterwordspacing

\bibitem{snyder2018x}
D.~Snyder, D.~Garcia-Romero, G.~Sell, D.~Povey, and S.~Khudanpur, ``X-vectors:
  Robust {DNN} embeddings for speaker recognition,'' in \emph{2018 IEEE
  International Conference on Acoustics, Speech and Signal Processing
  (ICASSP)}, 2018, pp. 5329--5333.

\bibitem{di2019mirage}
D.~Di~Carlo, A.~Deleforge, and N.~Bertin, ``{MIRAGE}: {2D} source localization
  using microphone pair augmentation with echoes,'' in \emph{2019 IEEE
  International Conference on Acoustics, Speech and Signal Processing
  (ICASSP)}, 2019, pp. 775--779.

\bibitem{dechorate}
\BIBentryALTinterwordspacing
D.~Di~Carlo, P.~Tandeitnik, C.~Foy, N.~Bertin, A.~Deleforge, and S.~Gannot,
  ``{dEchorate}: a calibrated room impulse response database for echo-aware
  signal processing,'' 2021, (under review). [Online]. Available:
  \url{https://hal.archives-ouvertes.fr/hal-03207860}
\BIBentrySTDinterwordspacing

\end{thebibliography}
%
%
%
%
%
%
%
%
%

\end{sloppy}
\end{document}